\documentclass[apj]{emulateapj}
\usepackage{times}
\usepackage[normalem]{ulem}
\usepackage{epsfig}
\usepackage{natbib}
\usepackage{amsfonts}
\usepackage{amsmath}
\usepackage{multirow}
\usepackage{enumerate}
\usepackage{verbatim}
\usepackage[usenames]{color}
\usepackage[plainpages=false, colorlinks=true, anchorcolor=blue, linkcolor=blue, citecolor=blue, bookmarks=false]{hyperref}
\newcommand{\be}{\begin{eqnarray}}
\newcommand{\ee}{\end{eqnarray}}
\newcommand{\lp}{\left(}
\newcommand{\rp}{\right)}

\newcommand{\slugcom}{Accepted for publication in The Astrophysical Journal}
\slugcomment{\slugcom}

\begin{document}

\normalsize


\title{Exploring the Potential Diversity of  Early Type Ia Supernova Light Curves}

\author{Anthony L. Piro\altaffilmark{1} and Viktoriya S. Morozova\altaffilmark{2}}

\altaffiltext{1}{Carnegie Observatories, 813 Santa Barbara Street, Pasadena, CA 91101, USA; piro@obs.carnegiescience.edu}

\altaffiltext{2}{Theoretical Astrophysics, California Institute of Technology, 1200 E California Blvd., M/C 350-17, Pasadena, CA 91125, USA}

\begin{abstract}
During the first several days after explosion, Type Ia supernova light curves probe the outer layers of the exploding star and therefore provide important clues for identifying their progenitors. We investigate how both the shallow $^{56}$Ni distribution and the presence of circumstellar material shape these early light curves. This is performed using a series of numerical experiments with parameterized properties for systematic exploration. Although not all of the considered models may be realized in nature (and indeed there are arguments why some of them should not occur), the spirit of this work is to provide a broader exploration of the diversity of possibilities. We find that shallower $^{56}$Ni leads to steeper, bluer light curves. Differences in the shape of the rise can introduce errors in estimating the explosion time and thus impact efforts to infer upper limits on the progenitor or companion radius from a lack of observed shock cooling emission. Circumstellar material can lead to significant luminosity during the first few days, but its presence can be difficult to identify depending on the degree of nickel mixing. In some cases, the hot emission of circumstellar material may even lead to a signature similar to interaction with a companion, and thus in the future additional diagnostics should be gathered for properly assessing early light curves.
\end{abstract}


\section{Introduction}

Although Type Ia supernovae (SNe Ia) are fundamentally important to many areas of astronomy and astrophysics, their actual progenitors are still uncertain. There remains a wide range of possible ways of exploding a white dwarf (WD), including stable accretion in a non-degenerate binary \citep{Whelan73}, mergers of C/O WDs \citep{Iben84,Webbink84}, accretion and detonation of helium shells on C/O WDs \citep{WoosleyWeaver94,Livne95}, and direct collisions \citep{Rosswog09,Raskin10,Thompson11,Kushnir13}. Even among these broad classes of explosion scenarios there are important differences, such as whether the merger ignition is triggered by a detonation in an accretion stream \citep{Guillochon10,Dan12}, in a violent merger involving massive WD \citep{Pakmor12}, or after a more long-term evolution of the merger remnant \citep[although this is generally viewed as unlikely, see][and the discussion below]{Shen12}. Understanding which of these scenarios are most prevalent and in what proportion remains an outstanding problem. Early SN Ia light curves hold promise for helping to distinguish between different progenitors with unique information, such as the shock cooling of the WD surface \citep{Piro10,Rabinak12}, collision of the explosion with the WD's companion \citep{Kasen10}, and the shallow profiles of velocity and radioactive nickel \citep{Piro12,Piro14,Mazzali15}.

Correctly interpreting the features seen during these early phases requires understanding what uncertainties may be present and how much diversity is possible. For example, placing limits on the radius of a possible companion star from the non-detection of a collision signature can depend strongly on the explosion time constraints \citep[see the discussion in][]{Shappee15}. The problem is that the explosion time can be inferred incorrectly from a simple light curve extrapolation if the SN has a dark phase \citep{Piro13}. In other cases, the early color evolution or detailed shape of a light curve may argue for interesting implications, such as the interaction with a companion \citep{Cao15,Liu15,Im15,Marion15}. The question is whether such interpretations are indeed unique or if there are other possible explanations. Furthermore, as more early observations of SN Ia light curves are collected, a range of rise times and rise shapes are seen \citep[][and references therein]{Firth15}. How to interpret this range and what it means physically for the SNe are poorly understood.

With these issues in mind we set out to explore the potential diversity of early SN Ia light curves. This is investigated using a variety of parameterized models that vary the $^{56}$Ni distribution along with the amount and extent of circumstellar material. The nickel distribution is explored because there is considerable diversity in how nickel is mixed in various models, from very stratified, centrally-ignited explosions to messy mergers. Circumstellar material is explored because it seems plausible that mass should be present around exploding WDs, at least in small amounts, given that accretion or merger is an integral part of any SN Ia scenario. In the end, it may be that many of the specific models presented here are not exactly realized in nature. The aim is that by exploring this potential diversity, important trends can be identified that will help with understanding what we can learn from observations independent of the exact models. Knowledge about the characteristic timescales and amplitudes of these features will assist in planning future transient surveys that will investigate these effects.

In Section \ref{sec:setup}, we describe the numerical setup we employed for generating and studying early SN Ia light curves. In Section \ref{sec:nickel}, we summarize our results from varying the distribution of $^{56}$Ni. This is followed up by exploring how the light curves are impacted by material around the exploding WD in Section \ref{sec:csm}, for which we again consider different levels of $^{56}$Ni mixing. We conclude in Section \ref{sec:conclusions} with a summary of our main results and a discussion of future work.

\section{Explosion and Light Curve Implementation}
\label{sec:setup}

We begin by describing our methods for generating background models, exploding these models, and then calculating the resulting light curves. For all of the work described below, we start with a $1.25\,M_\odot$ carbon/oxygen, degenerate core that was generated using the \texttt{MESA} stellar evolution code \citep{Paxton11}. This model is chosen since it is more massive than a typical WD, but also low mass enough that additional mass can be added in circumstellar material without being extremely super-Chandrasekhar. The main goal is to have a WD with an outer density profile that is largely dominated by degenerate electrons, since this is the region we will be probing with our light curves, and this region's profile will be somewhat insensitive to the exact mass of the star.

With these models in hand, we explode them and follow the propagation of the shock wave and the resulting light curves using the SuperNova Explosion Code \citep[\texttt{SNEC},][]{Morozova15}. Since \texttt{SNEC} was written with a focus on core-collapse SNe, its implementation for thermonuclear explosions requires some extra description, including what approximations must be made and what limitations and caveats this correspondingly places on the results presented here.

Both thermal bomb and piston-driven explosions are available in \texttt{SNEC}. For a WD, the thermal bomb explosions have difficulty in driving the overpressure necessary to generate a shock wave because of the large Fermi energy of the degenerate electrons. Thus, we focus on a piston-driven explosion here. For a velocity of $3.9\times10^9\,{\rm cm\,s^{-1}}$ placed on the inner cell for a timescale of $10^{-2}\,{\rm s}$, we are able to robustly get a $\approx1.1\times10^{51}\,{\rm erg}$ explosion, which we take as our fiducial model. Of course a real SN Ia is a thermonuclear explosion that burns as some combination of detonation and deflagration, which will impart a particular velocity distribution to the exploding material. At sufficiently low densities the burning wave will transition into a shock \citep{Piro10}, so the velocity profiles we find here using a purely shock driven explosion are more accurate at shallow depths \citep[although there still may be important non-spherical effects we are missing, e.g.,][]{Matzner13}. This is one of the reasons why we focus on early times in the light curves ($\lesssim8\,{\rm days}$ following explosion), since as the light curve probes closer to the center of the star our simulations are surely inaccurate in reproducing the velocity profiles that are present for realistic thermonuclear explosions.

The lack of a realistic explosion treatment means that we also do not self-consistently calculate the distribution of radioactive $^{56}$Ni deposited throughout the ejecta. Instead the $^{56}$Ni distribution must be placed by hand. This is not a problem for the present work since the $^{56}$Ni distribution is one of the key factors which we would like to explore and so having maximum flexibility in where and how much $^{56}$Ni is present is crucial. The actual $^{56}$Ni distributions used will be described in more detail when the results of the models are presented.

Another important aspect of these calculations is the opacity. \texttt{SNEC} currently has the ability to do flux-limited diffusion in thermodynamic equilibrium with a gray opacity. This is missing many important details required for a full treatment of the radiative transfer \citep[e.g.,][]{Dessart14}. For the heavy elements present in SNe Ia, the dominant opacity arises from line transitions mainly concentrated in the ultraviolet, which \texttt{SNEC} cannot follow. Instead we use the OPAL tabulated opacity tables for a carbon and oxygen rich mixture \citep{Iglesias96}. The OPAL opacities are obtained by solving for the occupation numbers of elements starting from the grand canonical ensemble of a system of electrons and nuclei interacting through the Coulomb potential \citep{Rogers92}. The opacities include contributions from the atomic lines of 21 elements (19 metals, including Fe and Ni), taking into account line broadening due to the Doppler, natural width, electron impacts, as well as scattering from neutral H and He. In the low-temperature regime (between $10^{2.7}\,{\rm K}$ and $10^{4.5}\,{\rm K}$), we use the tables of \citet{Ferguson05}, which, in addition to the atomic lines, take into account a wide range of molecular lines and the opacity due to the dust particles. The caveat of using these tables for the SN light curve calculations, especially for the large mass fractions of radioactive $^{56}$Ni, is that they assume local thermodynamical equilibrium and do not take into account non-thermal ionization and excitation by gamma-rays. This, and the fact that \texttt{SNEC} itself is based on the assumption of local thermodynamical equilibrium, smears out the impact of the atomic and molecular lines on the color evolution of the light curve. Also note that we have not included an ``opacity floor'' in our runs, even though this is typically included in the default version of \texttt{SNEC}. For SNe IIP, this is a standard method for correcting the differences between a Rosseland mean opacity and more detailed treatments \citep[see][and references therein]{Shigeyama90,Bersten11}. Since there is not precedent for this for SN~Ia modeling, we do not include it here.

This opacity setup reproduces the broad features that when the material is hot and/or dense the opacity is high, while as the material gets cold and/or diffuse the material is more transparent. Due to the heating from $^{56}$Ni, the opacity therefore ends up naturally high in the correct regions with a value roughly equal to the electron scattering component. This is actually not too different from the Rosseland mean values that \citet{Pinto00} found in their detailed summary of the opacity in iron-peak-dominated ejecta. Thus, although it might not be the correct physical reason, the opacities appear to be roughly correct values in both transparent and opaque regions. Nevertheless, this should still be tested by more detailed treatments of radiative transfer, especially since we do not have any wavelength dependent opacities. Hopefully our initial work here provides useful guidance for the trends expected in future more detailed work.

Finally, a critical limitation of using \texttt{SNEC} is that the calculations are one-dimensional. In contrast, many of the physical situations considered here will be demonstratively three-dimensional. For example, the $^{56}$Ni distribution for a violent merger will be concentrated in different regions, or the extended mass distribution in a collision will be highly aspherical. Therefore, many of these calculations will be more representative of some sort of average case with potential variations around our results depending on viewing angle.

\section{Nickel Distribution Impact}
\label{sec:nickel}

\begin{figure}
\epsscale{1.2}
\plotone{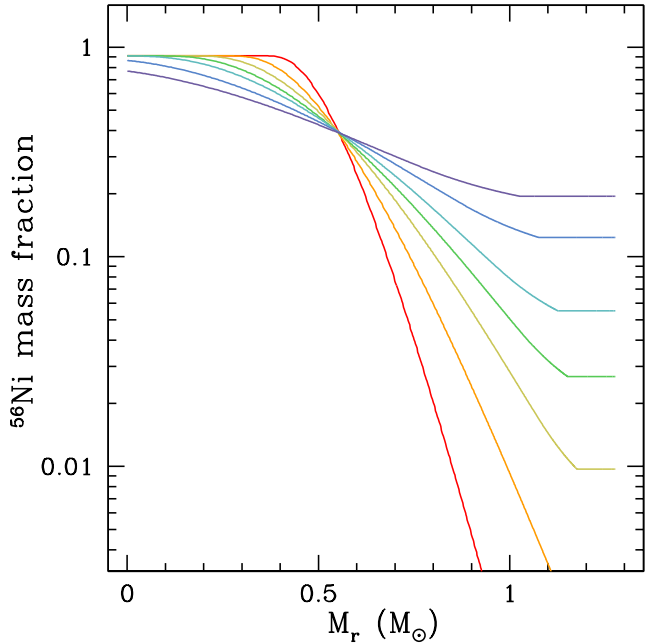}
\caption{Profiles of the mass fraction $^{56}$Ni as a function of the mass coordinate in the WD for the various levels of mixing considered in this work. These lines are produced using a boxcar averaging routine with widths of $0.05$, $0.075$, $0.1$, $0.125$, $0.15$, $0.2$, and $0.25\,M_\odot$ (from red to purple). }
\label{fig:ni}
\epsscale{1.0}
\end{figure}

Beginning with the bare WD described above, we now explore the impact of varying the distribution of $^{56}$Ni. It is well known that the total amount $^{56}$Ni sets the peak luminosity of SNe Ia, so we use a fixed amount of $^{56}$Ni of $0.5\,M_\odot$ \citep[similar to as observed,][]{Piro14b} and vary its distribution. This is performed with a ``boxcar'' averaging, as used, for example, in \citet{Kasen09} and \citet{Dessart12,Dessart13} to simulate mixing in the context of core-collapse SNe. We run a boxcar with a width from $0.05-0.25\,M_{\odot}$ through the model four times until we obtain a smooth profile (details of this procedure are available in the notes available at the \texttt{SNEC} website\footnote{\url{http://stellarcollapse.org/SNEC}}). These $^{56}$Ni distributions are presented in Figure \ref{fig:ni}.

\begin{figure}
\epsscale{1.2}
\plotone{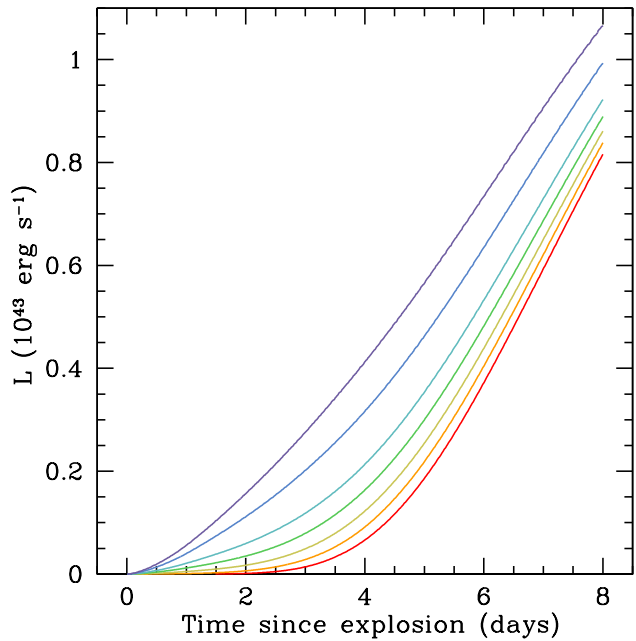}
\caption{Bolometric luminosity as a function of time during the first 8~days following explosion for a bare WD model. The line colors indicate the level of $^{56}$Ni mixing, which correspond to the profiles shown in Figure \ref{fig:ni}.}
\label{fig:linear_luminosity}
\epsscale{1.0}
\end{figure}

We then explode the WD model with each of these $^{56}$Ni distributions using the piston-driven setup described in Section \ref{sec:setup}. The resulting bolometric light curves over the first 8 days following explosion are presented in Figure \ref{fig:linear_luminosity}. For the lowest level of mixing (red curve), the rise is rather flat for the first $\sim3\,{\rm days}$, and then the light curve begins to rise more quickly. The entire luminosity seen in Figure \ref{fig:linear_luminosity} is dominated by $^{56}$Ni powering. Although shock heating is important at initial times, on the linear scale presented in this plot, its impact is negligible and thus it is not contributing to this change in shape of the light curve. Instead, when $^{56}$Ni is not highly mixed, the outer layers get very cold as they adiabatically expand at early times. This causes the photosphere to move in more quickly to where the shallowest regions of where $^{56}$Ni heating is just able to reach. This provides the early light curve. Then, at $\sim3\,{\rm days}$ for the red curve, the diffusion depth is really able to reach the $^{56}$Ni heating in earnest. This is when the light curve rises more dramatically. For the most mixed models shown in blue or purple (the curves further to the left in Figure \ref{fig:linear_luminosity}), the $^{56}$Ni is so shallowly mixed that the light curve always has this more rapid rise.

\begin{figure}
\epsscale{1.2}
\plotone{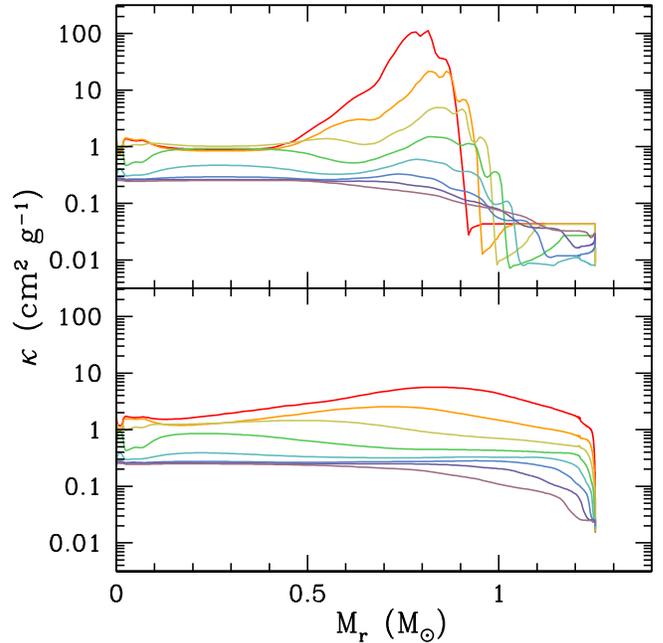}
\caption{Profiles of opacity as a function of mass coordinate for different times during the explosion, at snapshots of 0.1, 0.18, 0.32, 0.56, 1,  1.8, 3.2, and 5.6 days (from red to purple). The upper panel is the least mixed model considered (the red curves in Figures \ref{fig:ni} and \ref{fig:linear_luminosity} with a boxcar width of $0.05\,M_\odot$), while the bottom panel is the most mixed model (the purple curves in Figures \ref{fig:ni} and \ref{fig:linear_luminosity} with a boxcar with of $0.25\,M_\odot$).}
\label{fig:opacity}
\epsscale{1.0}
\end{figure}

\begin{figure}
\epsscale{1.2}
\plotone{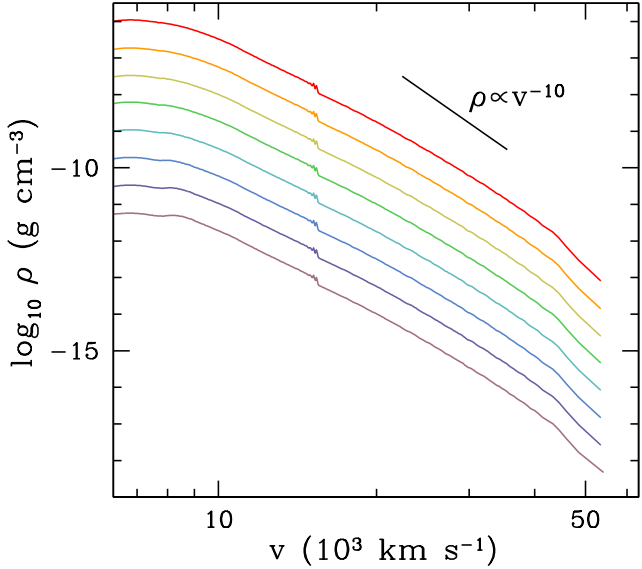}
\caption{Profiles of density as a function of velocity for different times during the explosion at the same snapshots as shown in the Figure \ref{fig:opacity}. We show an example power law of $\rho\propto v^{-10}$, as might be expected for shock acceleration in the decreasing density profile of a radiative star, demonstrating that the degenerate conditions here give an answer that is not too far different.}
\label{fig:rho}
\epsscale{1.0}
\end{figure}

The impact of the physics involved is clearly seen in Figure \ref{fig:opacity}, which shows the opacity as a function of depth at different snapshots spaced logarithmically in time from $0.1\,{\rm days}$ (red curves) to $5.6\,{\rm days}$ (purple curve). The top panel shows the least mixed model (the red curves in Figures \ref{fig:ni} and \ref{fig:linear_luminosity}). The opacity is relatively low in the shallower nickel-poor regions that get cold at early times. Then, as the material expands and nickel heating can spread, the opacity increases and flattens across the star. In contrast, the bottom panel of Figure \ref{fig:opacity}, shows the most mixed model (the purple curves in Figures \ref{fig:ni} and \ref{fig:linear_luminosity}). Here the opacity is  higher throughout the model, which keeps the photosphere and diffusion depths pushed out further. By $5.6\,{\rm days}$ (the purple curves in both panels), the opacity profiles in both the models actually look fairly similar, since now the shallow layers are getting heated by $^{56}$Ni in roughly the same way. Also, note that for the first $\sim0.3\,{\rm days}$, the opacity is rather flat near the surface of the star for the weakly mixed model. This is because the density is too high and the temperature too low for the opacity tables we utilize. Despite this difficulty (which only arises for the least mixed model), we tried increasing the outer opacity by a factor of 2 and decreasing it by a factor of 10 with no noticeable impact on the $V$-band light curve when the SN was brighter than an absolute magnitude of $-4$.

\begin{figure}
\epsscale{1.2}
\plotone{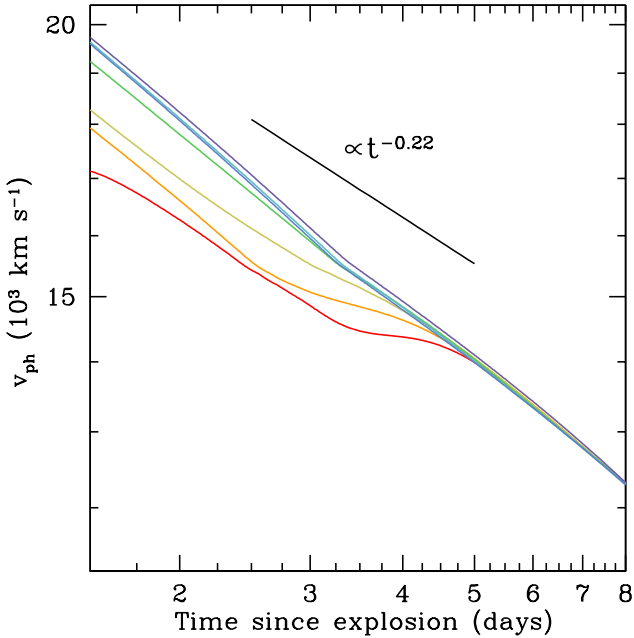}
\caption{Photospheric velocity $v_{\rm ph}$ as a function of time for the same models presented in Figure \ref{fig:linear_luminosity}. Note the logarithmic axis used to highlight any power law dependences.}
\label{fig:vphoto_bare_talk}
\epsscale{1.0}
\end{figure}

The photospheric velocity $v_{\rm ph}$ is also a helpful indicator of what depth in the star is being probed, since the shallower layers generally move faster than deeper layers due to shock acceleration near the surface of the star. This is shown in Figure \ref{fig:rho}, which presents the velocity profile at various snapshots in time. This demonstrates that the velocity profile we find is not all that different than what would be found for a radiative star, which is $\rho\propto v^{-10}$ \citep{Svirski12}.  In Figure \ref{fig:vphoto_bare_talk}, we summarize the time-dependent photospheric velocity for the same set of models with different levels of $^{56}$Ni mixing. For the least amount of mixing (the red curve), $v_{\rm ph}$ is lower at early times, since the transparency of the shallow, cool material allows deeper regions of the star to be probed earlier. In contrast, the most mixed model (purple curve) has a high velocity at early times because the hot, opaque material pushed the photosphere out further. At later times, when the opacities are roughly similar as shown in Figure \ref{fig:opacity}, the velocities also closely match independent of the level of mixing. As a comparison, a $v_{\rm ph}\propto t^{-0.22}$ power law is also denoted, as was derived by \citep{Piro14}. This appears to be too shallow at late times, and at early times there is considerable diversity. This may make using such a power law to constrain the explosion time difficult. The steeper drop of the velocity in the numerical models is due to the drop in opacity with time, which allows the photosphere to move into slower material more quickly. In contrast, the analytic estimate assumed a constant opacity in time and with depth. Note that although there is a density discontinuity apparent in Figure \ref{fig:rho} due to burning layers in the accreting WD model, this does not imprint itself onto the photospheric velocities in Figure \ref{fig:vphoto_bare_talk}.

\begin{figure}
\epsscale{1.2}
\plotone{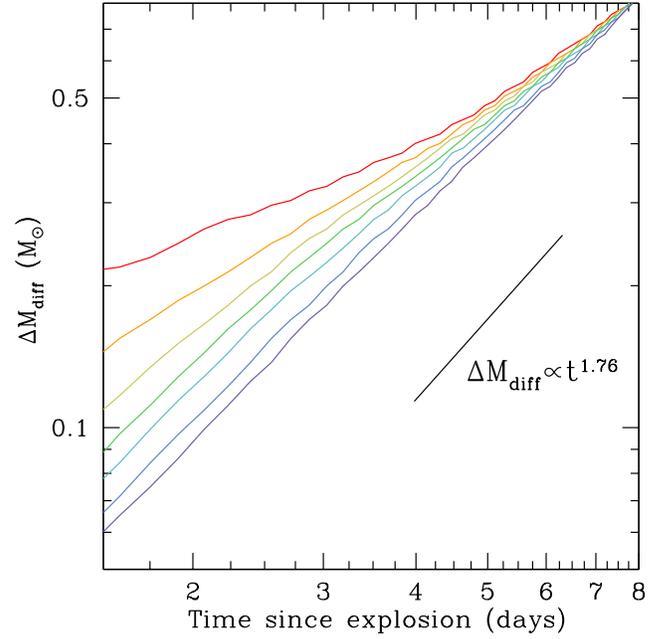}
\caption{Diffusion depth $\Delta M_{\rm diff}$ as a function of time for the same models presented in Figure \ref{fig:linear_luminosity}.}
\label{fig:tdiff}
\epsscale{1.0}
\end{figure}

A more direct way to see which depth of the star are being probed by the luminosity is the diffusion depth $\Delta M_{\rm diff}$. This is not the photospheric depth, but rather the depth at which the time for photons to diffuse out is roughly equal to the time since the explosion began \citep[see the more detailed discussion in]{Piro10}. This is plotted in Figure \ref{fig:tdiff} as a function of time for the various mixed models. In least mixed case (red), the $\Delta M_{\rm diff}$ is largest at early times, showing that this model probes deeper into the star than the more mixed models (purple). The power law $\Delta M_{\rm diff}\propto t^{1.76}$ was derived analytically in \citet{Piro14} and appears to accurate represent the simulations when nickel is dominating the heating. The main problem with the analytic relation is its normalization, which is roughly $25\%$ too shallow in comparison to these simulations. This means that nickel will not be quite as shallow as previously inferred by fitting rising light curves.

\begin{figure}
\epsscale{1.2}
\plotone{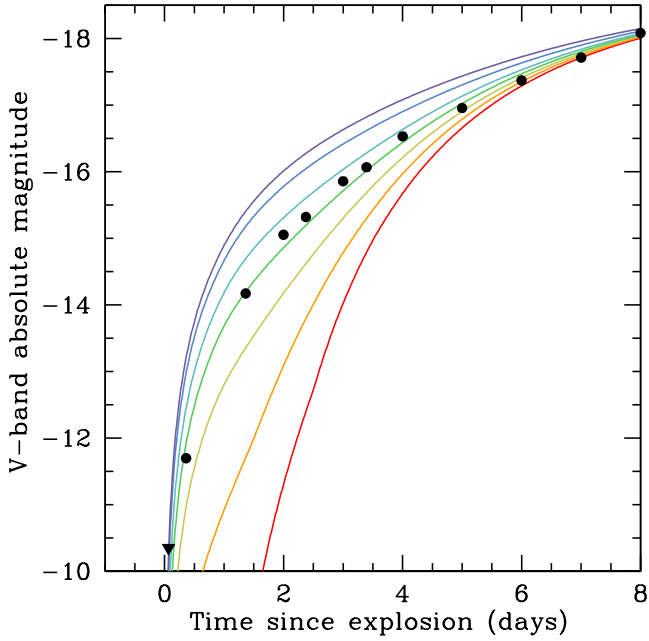}
\caption{Rising $V$-band light curves of the same set of models show in Figure \ref{fig:linear_luminosity}. A combination of $g$-band and $V$-band photometric measurements for SN 2011fe are also plotted as black points as an example early light curve (the first solid triangle is an upper limit). Note that in this band SN 2011fe is very close to a $t^2$ rise, so that this comparison demonstrates how much potential diversity there could be around $t^2$.}
\label{fig:photometry_11fe}
\epsscale{1.0}
\end{figure}

The comparison in Figure \ref{fig:linear_luminosity} highlights the difficulty in extrapolating a light curve to earlier times to infer the explosion time. When the mixing is strongest, the light curve extends relatively smoothly from times when early observations are often made at $\sim3-6\,{\rm days}$ back to the time of explosion. In contrast, the poorly mixed models show a strong inflection that could cause errors in inferring the explosion time. This difficulty is also seen in Figure \ref{fig:photometry_11fe}, where we present the $V$-band light curves for the same set of models as they would typically be presented observationally. For the lowest level of mixing (red curve), there can be a considerable dark phase \citep{Piro13} between actual moment of explosion and when the SN can first be detected depending on the depth of the observation. On the other hand, when the mixing is stronger (blue and purple lines) the light curves rise much more quickly.

As an example, we also include some $g$-band and $V$-band data points from SN 2011fe \citep{Nugent11,Vinko12} in Figure \ref{fig:photometry_11fe}. This is not meant to be a fit to the data, since the $^{56}$Ni distribution is varied in a completely artificial way.  Nevertheless, this shows that at least among these $^{56}$Ni distributions, SN 2011fe appears to be moderately mixed. This is consistent with the inferences of \citet{Piro12}, \citet{Piro14}, and \citet{Mazzali15}, although note that \citet{Dessart14b} find that some spectral properties of SN 2011fe may be difficult to explain with just mixing. If this theoretical curve is a good description for SN 2011fe, then it also argues that the dark phase for this SN is relatively short. This strengthens constraints on the progenitor and companion radii due to the lack of observed shock cooling emission \citep{Piro10,Kasen10,Bloom12}.

\begin{figure}
\epsscale{1.2}
\plotone{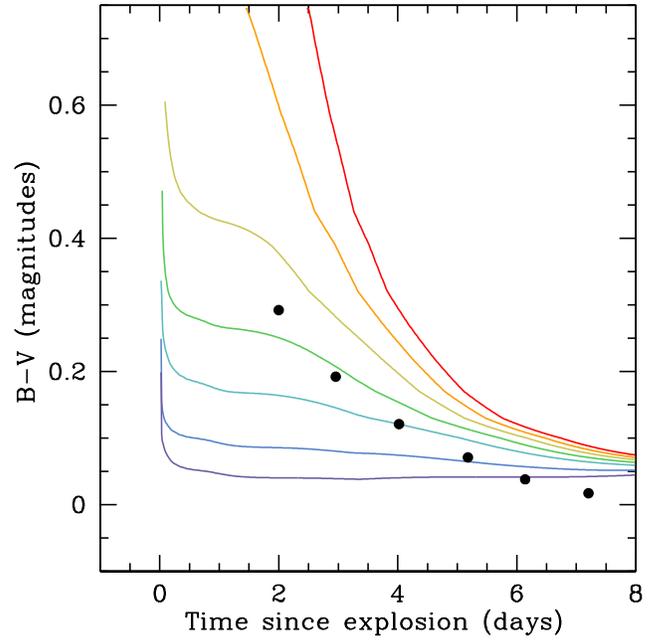}
\caption{The $B-V$ color evolution of the same set of models from Figure  \ref{fig:linear_luminosity} in comparison to the colors for SN 2011fe \citep{Pereira13}.}
\label{fig:color_bare}
\epsscale{1.0}
\end{figure}

A strong indication of the level of mixing is also the color evolution. This is highlighted in Figure \ref{fig:color_bare}, where we show the $B-V$ color for the same set of differently mixed models as in Figure \ref{fig:photometry_11fe}. The more highly mixed models are bluer at earlier time due to the shallow heating from $^{56}$Ni, while the less mixed models are significantly redder. These calculations come with a big caveat, namely we have not included the lines opacities expected from iron-group elements, which would impact the $B$-band magnitude and thus also these colors. Thus, this comparison is meant to be more qualitative than quantitative and mainly serves to demonstrate the general trend expect as the mixing is changed. Nevertheless, the moderately mixed model is again roughly similar to SN 2011fe, again indicating that this event is at least consistent with this level of mixing.

To summarize, just varying the distribution of $^{56}$Ni near the surface of an SN Ia can introduce considerable variation in the rise, with the strongest features being present in the first $\sim3\,{\rm days}$. This varies from steep quick rises when the $^{56}$Ni is shallow to shallower rises with a considerable dark phase when the $^{56}$Ni is deeper (Figure \ref{fig:photometry_11fe}). The transition from slow diffusion of $^{56}$Ni heating to more full-fledged heating can produce inflections in the light curve, which are especially apparent when the rise is plotted linearly rather than in magnitudes (see Figure \ref{fig:linear_luminosity}). Color evolution can be a useful discriminant, with shallower heating leading to bluer early-time emission.

\section{Circumstellar Material}
\label{sec:csm}

Most scenarios for producing SNe Ia involve some sort of mass transfer process, whether it be accretion from another star or a merger of two WDs.  These should in principle leave excess material around the exploding WD. This motivates a study of the impact of such material on the rising light curve, which may help discriminate between different explosion scenarios.

\subsection{Circumstellar Setup}

\begin{figure}
\epsscale{1.2}
\plotone{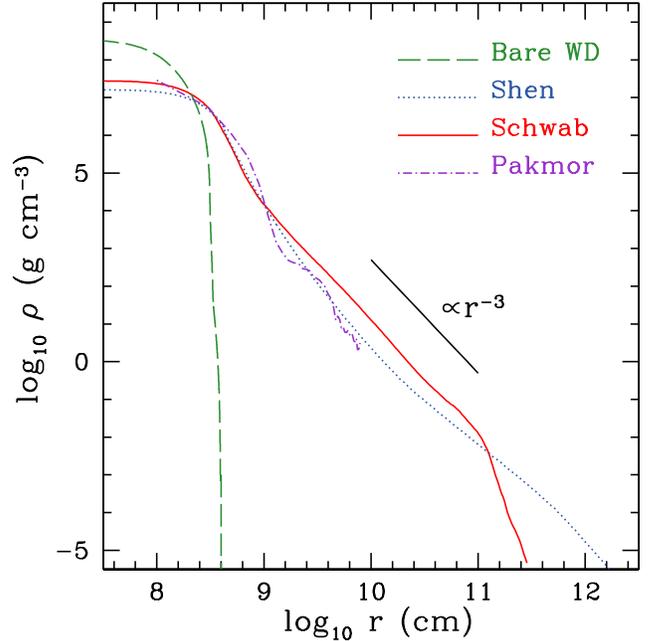}
\caption{Example density profiles showing a diversity of outcomes possible for the circumstellar material following a double WD merger. These profiles are a $1.25\,M_\odot$ WD (green, dashed line), equatorial material during a violent merger (\citealp{Pakmor12}, purple, dot-dashed line), the nearly spherical profile after the material has viscously relaxed (\citealp{Schwab12}, red, solid line), and the material after it has expanded to a red supergiant-like size (\citealp{Shen12}, blue, dotted line).}
\label{fig:density}
\epsscale{1.0}
\end{figure}

For some guidance in what sort of mass distributions to consider, we turn to studies of post merger density distributions as presented by \citet{Pakmor12}, \citet{Schwab12}, and \citet{Shen12}, and summarized in Figure \ref{fig:density}. These calculations roughly show how circumstellar material  evolves from immediately following the merger ($\sim$~seconds), to many viscous times ($\sim$~hours), to many thermal times ($\sim$~1000 yrs), respectively. Even over this large range of timescales, this comparison demonstrates that a $\rho\propto r^{-3}$ profile is a good description of all these density profiles. This is because the large heating near the base of the material leads to a roughly constant flux, radiative envelope with an $n=3$ polytropic index. Motivated by this, we consider similar density distributions for extended material in this paper of $\rho_{e}\propto r^{-3}$, so that the mass of the extended material is simply
\be
	M_e = 4\pi \int_{R_*}^{R_e} \rho_e(r)r^2dr \approx 4\pi r_e^3\rho_{e,0}\log(R_e/R_*),
\ee
where $R_*$ is the radius of the underlying WD and $\rho_{e,0}$ is a normalization constant for the density profile. Strictly speaking, the models that evolve for large periods of time following merger are not expected to necessarily make SNe Ia and instead lead to accretion induced collapse \citep{Saio98,Shen12,Schwab15}. Nevertheless, in the spirit of this work, we will still explore a variety of density distributions with different values of $M_e$ and $R_e$, even for cases that might not reach an SN Ia, to see what the corresponding observational signatures should be.

\begin{figure}
\epsscale{1.2}
\plotone{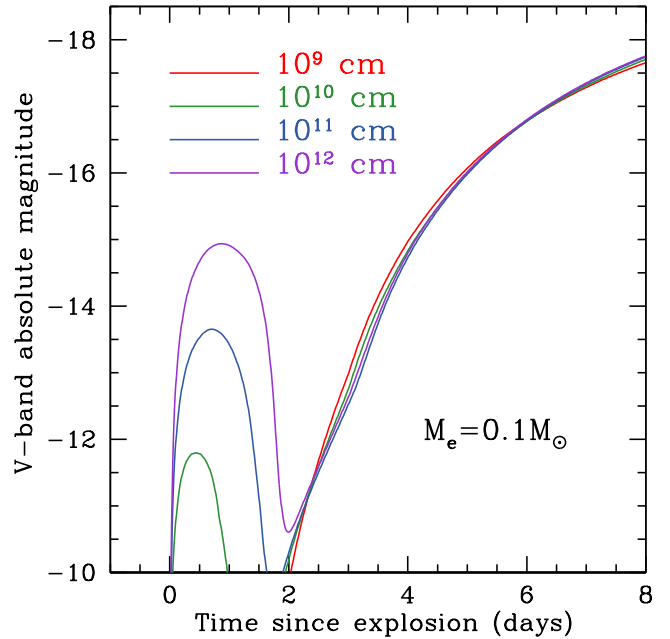}
\caption{Rising $V$-band light curves for models with $M_e=0.1\,M_\odot$ of extended material with a variety of radii from $10^9$ to $10^{12}\,{\rm cm}$, as indicated. Here, mixing is kept low with a boxcar width of $0.05\,M_\odot$.}
\label{fig:photometry_m0.1_s0.05}
\epsscale{1.0}
\end{figure}

\subsection{Extended Material Light Curves}

Our first set of circumstellar calculations are shown in Figure \ref{fig:photometry_m0.1_s0.05}. Here we set the external mass to $M_e=0.1\,M_\odot$ and vary the outer radius from $R_e=10^9\,{\rm cm}$ up to $10^{12}\,{\rm cm}$. The explosion is triggered in the same way as all previous models, and the $^{56}$Ni mixing is kept low with a boxcar width of $0.05\,M_\odot$, the same as the red curves in Figures \ref{fig:ni}, \ref{fig:linear_luminosity}, \ref{fig:vphoto_bare_talk}, \ref{fig:photometry_11fe}, and \ref{fig:color_bare}. The impact of the extended material on the light curve is dramatic, resulting in a bright first peak from the shock cooling of this extended material. This is reminiscent of the double-peaked light curves of some SNe IIb, which have also been attributed to extended material around the progenitor star \citep{Bersten12,Nakar14}. Furthermore, the roughly parabolic shape is similar to the analytic light curve expected for this material \citep{Piro15}. The observation of such a feature would be extremely useful for constraining the properties of extended material. The peak luminosity scales roughly proportional to $R_e$ and the width of the first peak would scale roughly as $M_e^{1/2}$. Unfortunately, an early peak like this has never been observed (although admittedly only a few SNe Ia have the time coverage and depth necessary to rule out the presence of such a feature).

\begin{figure}
\epsscale{1.2}
\plotone{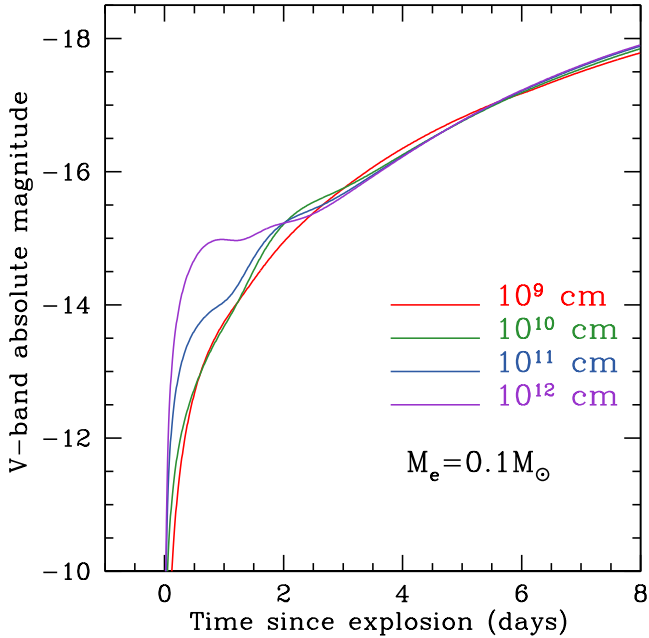}
\caption{The same as Figure \ref{fig:photometry_m0.1_s0.05}, but with increased $^{56}$Ni mixing using a boxcar width of $0.15\,M_\odot$. This highlights that a time coverage of $\lesssim1\,{\rm day}$ is required to really identify these features.
}
\label{fig:photometry_m0.1_s0.1}
\epsscale{1.0}
\end{figure}

Partially motivated by the lack of an observed double-peaked SN Ia, we next modify the previous calculation by increasing the amount of $^{56}$Ni mixing. We now use a boxcar width of $0.15\,M_\odot$, as for the cyan curves in Figures \ref{fig:ni}, \ref{fig:linear_luminosity}, \ref{fig:vphoto_bare_talk}, \ref{fig:photometry_11fe}, and \ref{fig:color_bare}. Physically, this could correspond to an asymmetric explosion where significant $^{56}$Ni was generated in outer regions or even an explosion with significant instabilities and mixing. The resulting $V$-band light curves are summarized in Figure \ref{fig:photometry_m0.1_s0.1}. The mixing causes the first peak to be much less prominent, although note that the contribution shock from cooling emission is roughly the same as in Figure \ref{fig:photometry_m0.1_s0.05}.

\begin{figure}
\epsscale{1.2}
\plotone{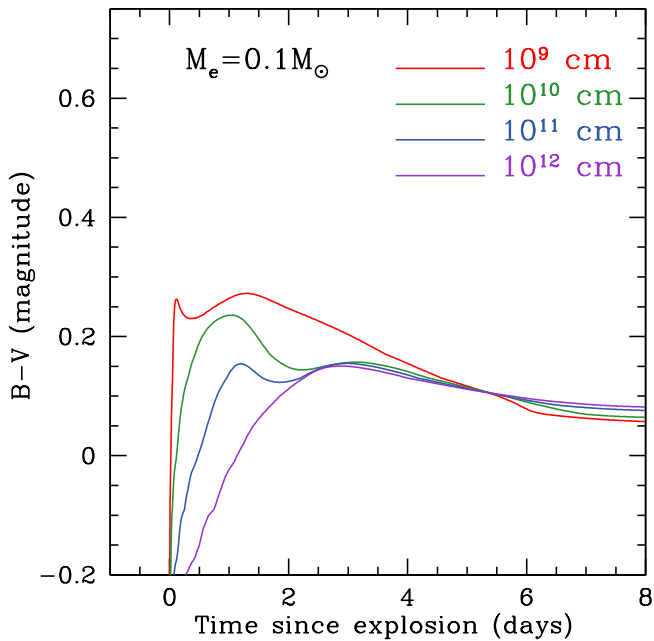}
\caption{The $B-V$ color evolution of the models shown in Figure \ref{fig:photometry_m0.1_s0.1}. Extended material results in significantly bluer colors at early times, an effect which is stronger with larger $R_e$.}
\label{fig:colors_m0.1_s0.1}
\epsscale{1.0}
\end{figure}

In principle, mixing like this could help hide the presence of extended material around an SN Ia. In the rising light curve of ASASSN-14lp  \citep{Shappee15}, the first data at roughly $\sim1.5\,{\rm days}$ past explosion is slightly raised with respect to what one might expect for a smooth rise. This is enticingly similar to the effect of extended material shown in Figure \ref{fig:photometry_m0.1_s0.1}. Unfortunately, because the data was too sparse (although well-sampled in comparison to most any other early SN Ia observations!) it could not be concluded whether or not the rise of ASASSN-14lp was exemplary or not. In the future, our work here will hopefully help motivate sub-day timescale observations during the first $\sim3\,{\rm days}$ of SNe Ia to really nail down whether processes like this are in fact occurring.

As discussed before for the $^{56}$Ni distribution, color evolution may be another important discriminant for interpreting early light curves. Thus we plot the $B-V$ color evolution for the models from Figure \ref{fig:photometry_m0.1_s0.1} in Figure \ref{fig:colors_m0.1_s0.1}. Even when the changes to the photometric light curves are relatively small, much more pronounced differences can be seen in the colors. This is because extended material with a larger $R_e$ suffers from relatively less adiabatic cooling after a fixed set of time in comparison to a smaller $R_e$, which expands much more in comparison to its initial radius. The result is that a larger $R_e$ has much bluer colors. This difference ends at approximately $\sim3\,{\rm days}$, when the colors settle down to be roughly the same independent of the extended material radius.

Although this variety of early color evolution could make it an important probe of the extended material, it is likely not unique. In particular, the impact of the interaction with a companion \citep{Kasen10} could also produce strong color evolution where now bluer early colors indicate a larger radius companion \citep{Marion15}. Therefore, it will be important to identify other features that can discriminate between these scenarios of extended material versus a companion. One way would be looking for the X-rays expected to be associated with shock interaction with the companion \citep{Kasen10}. Another way would be searching for signs of hydrogen in the late time spectra from material stripped from the companion \citep[e.g.,][]{Mattila05,Leonard07,Shappee13,Lundqvist15}.

\begin{figure}
\epsscale{1.2}
\plotone{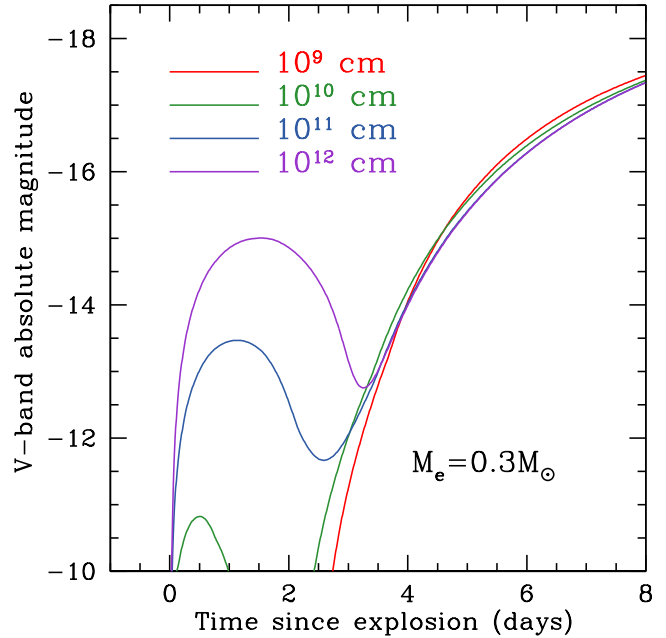}
\caption{The same as Figure \ref{fig:photometry_m0.1_s0.05}, but with an increased extended material mass of $M_e=0.3\,M_\odot$. The mixing is kept low with a boxcar width of $0.05\,M_\odot$.}
\label{fig:photometry_m0.3_s0.05}
\epsscale{1.0}
\end{figure}

\begin{figure}
\epsscale{1.2}
\plotone{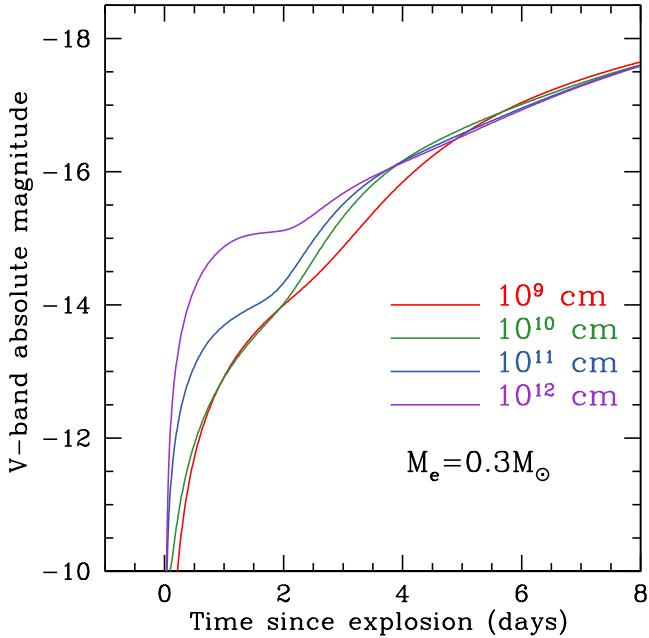}
\caption{The same as Figure \ref{fig:photometry_m0.3_s0.05}, but with an increased $^{56}$Ni mixing using a boxcar width of $0.15\,M_\odot$.}
\label{fig:photometry_m0.3_s0.15}
\epsscale{1.0}
\end{figure}

\begin{figure}
\epsscale{1.2}
\plotone{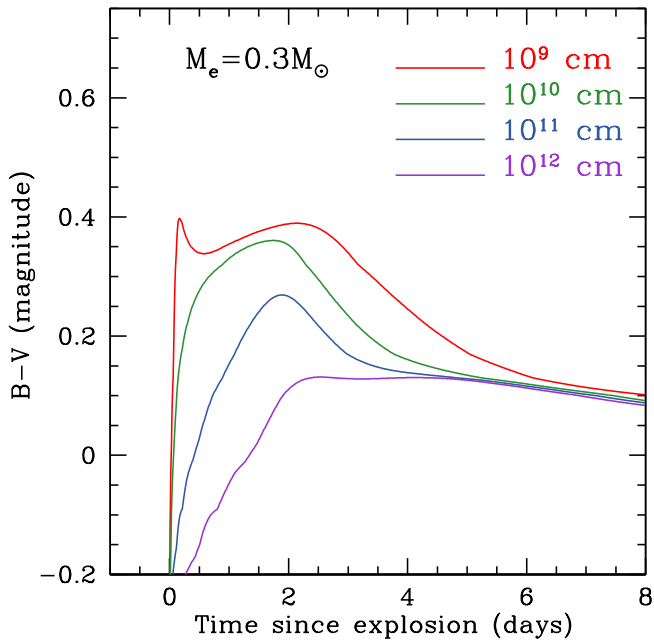}
\caption{The $B-V$ color evolution of the models shown in Figure \ref{fig:photometry_m0.3_s0.15}.}
\label{fig:colors_m0.3_s0.15}
\epsscale{1.0}
\end{figure}

Just to explore slightly more parameter space, we also consider an extended mass of $M_e=0.3\,M_\odot$ with low and high $^{56}$Ni mixing in Figures \ref{fig:photometry_m0.3_s0.05} and \ref{fig:photometry_m0.3_s0.15}, respectively. Not surprisingly, the width of the first feature increases approximately as $M_e^{1/2}$ as expected for the diffusion timescale, while the peak luminosities are roughly the similar for the same $R_e$ as considered before. The largest qualitative difference is in the light curve morphology for high mixing and relatively small extended material (the red curve in Figure \ref{fig:photometry_m0.3_s0.15}). Here, the light curve shows an inflection due to the $^{56}$Ni being mixed up into the relatively low density extended material (rather than this being due to shock cooling emission). As for the previous examples, we also include the color evolution in Figure \ref{fig:colors_m0.3_s0.15}, which shows the expected trends of bluer colors for more extended material.

\section{Conclusions and Discussion}
\label{sec:conclusions}

In this work we explored a variety of toy SN Ia models to theoretically investigate how changes in the $^{56}$Ni distribution and presence of circumstellar material may impact the rising light curve during the first $\sim8\,{\rm days}$. Our main conclusions are as follows.
\begin{itemize}
\item Models with more $^{56}$Ni mixed to shallower depth leads to a fast rise, a bluer early color, and less of a dark phase following the actual time of explosion.
\item Low $^{56}$Ni mixing may lead to a dark phase that can cause problems when light curves are extrapolated back in time to infer the moment of explosion. This will be quantified better in future work.
\item Extended material can lead to significant shock cooling emission, which scales roughly proportional to the radius of the material.
\item Depending on the level of mixing, the shock cooling emission can be more or less distinct from the main rise of the SN.
\item Looking at the color evolution can be useful for inferring shock cooling emission when it is not as apparent in the broad-band light curve.
\item This color can potentially evolve similarly to the interaction with a companion, so other diagnostics (such as the presence of hydrogen at late times) should be consulted to determine between companion and circumstellar material scenarios.
\end{itemize}
This work highlights the typical timescales and luminosities of the features associated to these effects. In particular, most of the features we explored showed the largest impact during the first $\sim3\,{\rm days}$ following explosion. Multiple observations will therefore be needed within this timeframe to test and measure these effects. The typical magnitudes during this time were $M_v \approx -12$ to $-15$. Furthermore, multiple bands (and probably bluer ones like $B$ and $V$ rather than $R$) are necessary for discerning properties like the color evolution.

There are a number of ways in which this work could be improved in the future. Chief among these is a more detailed treatment of the radiative transfer and the opacities. The dark phase we infer for the models with the deepest $^{56}$Ni distributions (see Figure \ref{fig:photometry_11fe}) depends how quickly the opacity decreases as the outer layers expand in cool and then how rapidly the opacity increases again once the $^{56}$Ni heating builds up. Both of these processes can be impacted by the details of opacity and radiative transfer, and this can also introduce different effects depending on the broad band filter of interest. This is especially true of the colors we infer in Figure \ref{fig:color_bare}. Although the models with shallower $^{56}$Ni are bluer at earlier times due to its heating, the large line opacities associated with iron-peak elements will have the opposite effect of increasing the opacity at these bluer wavelengths. Which effect wins in the end, and how it depends on the amount and distribution of $^{56}$Ni, should be explored in future work.

Another improvement that should be made in followup work is a better implementation of the explosion itself. Here we just used a shock since it was easy to control for our numerical experiments, and it should roughly have the correct properties near the surface of the star. This limited us to studying only the first $\sim8\,{\rm days}$ of these explosions. Nevertheless, in the deeper layers, a thermonuclear explosion will have a very different density and velocity profile than a shock, and will even depend on the details of how the burning proceed (detonation, deflagration, delayed detonation, gravitationally-confined-detonation, etc.). With a more complete treatment of the explosion for a number of different scenarios, connections can be made between the early features we identify here and the properties of the light curves at later times.

\acknowledgements
We thank the anonymous referee, Luc Dessart, Robert Firth, Ryan Foley, Benjamin Shappee, Jeffrey Silverman, and Mark Sullivan for helpful feedback, Ken Shen for providing the bare $1.25\,M_\odot$ WD model from \texttt{MESA} that was used for this work, and Ruediger Pakmor and Josiah Schwab, as well as Ken Shen again, for sharing their density profiles shown in Figure \ref{fig:density}. VSM is supported in part by the National Science Foundation under award Nos.\ AST-1205732 and AST-1212170, by Caltech, and by the Sherman Fairchild Foundation. Some computations were performed on the Caltech compute cluster Zwicky (NSF MRI-R2 award no.\ PHY-0960291).

\bibliographystyle{apj}

\end{document}